\documentclass[12pt,amsbsy]{article}
\newcommand{\beq}{\begin{equation}}
\newcommand{\eeq}{\end{equation}}
\newcommand{\beqn}{\begin{eqnarray}}
\newcommand{\eeqn}{\end{eqnarray}}

\newcommand{\de}{\mbox{${\delta}$}}

\begin{document}

\begin{titlepage}

\vspace{1cm}

\begin{center}
{\bf \large Upper limits on density of dark matter in Solar
system}
\end{center}

\vspace{5mm}

\begin{center}
I.B. Khriplovich \footnote{khriplovich@inp.nsk.su}\\
{\em Budker Institute of Nuclear Physics\\
11 Lavrentjev pr., 630090 Novosibirsk, Russia,\\
and Novosibirsk University}
\end{center}

\begin{center}
E.V. Pitjeva \footnote{evp@quasar.ipa.nw.ru}\\
{\em Institute of Applied Astronomy\\
10 Kutuzov Quay, 191187 St. Petersburg, Russia}
\end{center}
\bigskip

\begin{abstract}
The analysis of the observational data for the secular perihelion
precession of Mercury, Earth, and Mars, based on the EPM2004
ephemerides, results in new upper limits on density of dark matter
in the Solar system.

\end{abstract}


\end{titlepage}

According to present observational data (see, for instance,
\cite{ber}), the dark matter density in the region of the Milky
Way Galaxy, where our Solar system is situated, is
\begin{equation}\label{gal}
\rho_{\rm \;dm} \sim 0.5\times 10^{-24}\, {\rm g/cm^3}\,.
\end{equation}
However, not so much is known about the local dark matter
distribution inside our Solar system itself. A direct information
on this distribution, even rather modest one as compared to the
above number, is of certain interest.

Previous estimates of the dark matter density in the Solar system
[2\,--\,4] resulted in the upper limits on the level of
\begin{equation}\label{ag}
\rho_{\rm \;dm} < 10^{-16}\;{\rm g/cm^3}\,.
\end{equation}
The bounds obtained in \cite{an,and} were based on the
investigation of the possible effect of dark matter on the orbit
of Uranus. On the other hand, the limit derived in \cite{gr}
resulted from the analysis of the perihelion precession of the
asteroid Icarus.

Here we present much stronger upper limits on the dark matter
content in the Solar system following from analysis of the
perihelion precession of Mercury, Earth, and Mars. Our limits are
based on the precision EPM ephemerides constructed in \cite{pit1}
by simultaneous numerical integration of the equations of motion
for the nine major planets, the Sun, and the Moon in the
post-Newtonian approximation. Such subtle effects as the influence
of 301 large asteroids and of the ring of small asteroids, as well
as the solar oblateness, were included into the calculation. The
EPM ephemerides resulted from a least squares adjustment to more
than 317000 position observations (1913--2003) of different types,
including radiometric and optical astrometric observations of
spacecraft, planets, and their satellites.

The conclusions of the present paper are based on the possible
corrections to the secular perihelion precession $\de\phi$ (i.e.
on the deviations of the results of theoretical calculations from
the observational data) of three planets \cite{pit2} obtained from
about 250000 high-precision American and Russian ranging to
planets and spacecraft (1961-2003), including in particular
Viking-1,2, Pathfinder, Mars Global Syrveyor, and
Odyssey.\footnote{The corresponding correction from \cite{pit2} to
the perihelion precession of Venus is not used here, since it is
much larger and less accurate. The reasons are as follows. On the
one hand, Venus moves around the Sun more slowly than Mercury. On
the other hand, for Venus there are no high-precision ranging,
like those to martian landers or orbiting spacecraft, determining
the accuracy of the corrections for Mars and Earth.} They are
presented in Table 1.

\begin{center}
\begin{tabular}{|c|c|c|c|} \hline
 & & &  \\
&Mercury  & Earth & Mars \\
& & &  \\ \hline
& & &  \\
$''$ per century & $-$\,0.0036\,$\pm$\,0.0050 &
$-$\,0.0002\,$\pm$\,0.0004 &
0.0001\,$\pm$\,0.0005\\
& & &  \\ \hline
& & & \\
$(\de\phi/2\pi)\times
10^{11}$ & $-$\,0.67\,$\pm$\,0.93  & $-$\,0.15\,$\pm$\,0.31 &  0.14\,$\pm$\,0.73\\
 & & & \\ \hline
\end{tabular}

\vspace{5mm} Table 1. Corrections to secular perihelion precession
of planets
\end{center}

Let us address now the possible influence of dark matter on the
perihelion precession. To simplify our estimates, we will consider
the dark matter as nonrelativistic dust with spherically symmetric
density $\rho(r)$.

Then the correction $\de F(r)$ to the gravitational force acting
upon a planet with mass $m$ situated at a distance $r$ from the
Sun is found easily by means of the Gauss theorem:
\beq\label{def}
F(r) = -\, k \,m\,\frac{\mu(r)}{r^2}\,,
\eeq
where
\[
\mu(r) = 4\pi\int_0^r\rho(r_1)r_1^2 dr_1\,
\]
is the total mass of dark matter inside a sphere of radius $r$,
and $k$ is the Newton gravitational constant. The corresponding
correction to the gravitational potential is
\begin{equation}\label{deu}
\de U(r) = \, k \,m\,\int_0^r dr_2 \frac{\mu(r_2)}{r_2^2}\,.
\end{equation}
This correction shifts the perihelion of the planet orbit by angle
\begin{equation}
\de\phi = \,\frac{d}{dL}\,\frac{2m}{L}\,\int_0^\pi d\phi
\,r^2\,\de U(r)
\end{equation}
per period \cite{ll}; here $L$ is the planet angular momentum.
With $\de U(r)$ given by formula (\ref{deu}), and under the
assumption (made also in [2\,--\,4]) that $\rho_{\rm \;dm}$
remains constant at the distances discussed, we arrive after some
transformations at the result:
\begin{equation}
\frac{\de\phi}{2\pi} =
-\,\frac{3}{2}\,\frac{\mu(r)}{M_\odot}\sqrt{1-e^2}\,,
\end{equation}
where $M_\odot$ is the mass of the Sun and $e$ is the eccentricity
of the planet orbit. For Mercury, Earth, and Mars $e$ is small,
about 0.21, 0.02, and 0.09, respectively, and will be neglected in
the estimates below.

Now the results of \cite{pit2} presented in Table 1 can be
interpreted as upper limits on the mass of dark matter inside the
orbits of corresponding planets. Using the data in the last line
of Table 1, we formulate these upper limits as follows:
\begin{equation}\label{me}
\mu(0.39\; {\rm au}) < 6\times 10^{-12}\;M_\odot\,,
\end{equation}
\begin{equation}\label{e}
\mu(1\;{\rm au}) < 2\times 10^{-12}\;M_\odot\,,
\end{equation}
\begin{equation}\label{ma}
\mu(1.52\; {\rm au}) < 6\times 10^{-12}\;M_\odot\,;
\end{equation}
here we indicate in brackets the distance (in astronomical units,
au) from the Sun of the corresponding planet, Mercury, Earth, and
Mars, respectively.

At last, we convert these upper limits into the limits on the dark
matter density. Then, the Mercury limit (\ref{me}) results in the
bound
\begin{equation}
\rho_{\rm \;dm} < 10^{-17}\; {\rm g/cm^3}\,,
\end{equation}
somewhat better than (\ref{ag}).

The bounds following from the Earth and Mars limits, (\ref{e}) and
(\ref{ma}), are much stronger. They practically coincide and
constitute
\begin{equation}\label{re}
\rho_{\rm \;dm} < 3 \times 10^{-19}\; {\rm g/cm^3}\,.
\end{equation}
The result (\ref{re}), though being far from the global (Galaxy)
estimate (\ref{gal}), improves essentially the previous local
(Solar system) limits (\ref{ag}).

\begin{center}***\end{center}
We are grateful to A.E. Bondar, S.M. Kopeikin, A.A. Pomeransky,
V.G. Serbo, and M.I.~Vysotsky for the interest to the work, useful
discussions, and pointing out two errors in the earlier versions
of this note. The investigation was supported in part by the
Russian Foundation for Basic Research through Grant No.
05-02-16627.

\end{document}